\documentclass[]{spie}  

\usepackage{amsmath,amsfonts,amssymb}
\usepackage{graphicx}
\usepackage[colorlinks=true, allcolors=blue]{hyperref}

\title{Effects of Mirror Seeing on High-Contrast Adaptive Optics Instruments}
 
\author[a,*]{Melisa Tallis}
\author[a,b]{Vanessa P. Bailey}
\author[a]{Bruce Macintosh}
\author[c]{Lisa A. Poyneer}
\author[a]{Jean-Baptiste Ruffio}
\author[d]{Thomas L. Hayward}
\author[d]{Fredrik T. Rantakyr{\"o}}
\author[a,e]{Jeffrey K. Chilcote}
\author[f]{Dmitry Savransky}
\author[ ]{and the GPI Team.}

\affil[a]{Kavli Institute for Particle Astrophysics \& Cosmology, Physics Department, Stanford University, Stanford, CA, 94305, USA}
\affil[b]{Jet Propulsion Laboratory, California Institute of Technology, Pasadena, CA, 91109, USA}
\affil[c]{Lawrence Livermore National Laboratory, 7000 East Ave, Livermore, CA, 94550, USA}
\affil[d]{Gemini Observatory, Casilla 603, La Serena, Chile}
\affil[e]{Department of Physics, University of Notre Dame, Notre Dame, IN, 46556, USA}
\affil[f]{Sibley School of Mechanical and Aerospace Engineering, Cornell University, Ithaca, NY, 14853, USA}

\begin{document} 
\maketitle

\begin{abstract}
Ground-based direct imaging surveys like the Gemini Planet Imager Exoplanet Survey (GPIES) rely on Adaptive Optics (AO) systems to image and characterize exoplanets that are up to a million times fainter than their host stars. One factor that can reduce AO performance is turbulence induced by temperature differences in the instrument's immediate surroundings (e.g.: ``dome seeing'' or ``mirror seeing''). In this analysis we use science observations, AO telemetry, and environmental data from September 2014 to February 2017 of the GPIES campaign to quantify the effects of ``mirror seeing'' on the performance of the GPI instrument. We show that GPI performance is optimal when the primary mirror (M1) is in equilibrium with the outside air temperature. We then examine the characteristics of mirror seeing by calculating the power spectral densities (PSD) of spatial and temporal Fourier modes. Inside the inertial range of the PSDs, we find that the spatial PSD amplitude increases when M1 is out of equilibrium and that the integrated turbulence may exhibit deviations from Kolmogorov atmospheric turbulence models and from the 1-layer frozen flow model. We conclude with an assessment of the current temperature control and ventilation strategy at Gemini South.

\end{abstract}


{\noindent \footnotesize\textbf{*}Melisa Tallis,  mtallis@stanford.edu }
\section{Introduction}
\label{sect:intro}  
Direct imaging is a technique used for the detection and characterization of gas giant exoplanets and their formation environments. Each newly imaged planet helps constrain planet populations and provides a laboratory to study the formation, evolution, and atmospheric chemistry of massive, young, self-luminous planets, which are also referred to as ``young Jupiters.''\cite{Marois_2008, Nielsen_2014, Macintosh_2015, Bowler_2016, Rajan_2017, Keppler_2018, Stone_2018} Known young Jupiters are $\sim 10^{4}$--$10^{7}$ times fainter than their host stars in the near-infrared and are often separated by less than $0.5~''$. Specifically-designed high-contrast instrumentation on large ground-based telescopes can spatially resolve these faint sources from their host stars. The Gemini Planet Imager (GPI) on the 8~m telescope at the Gemini South Observatory (GS), is one such instrument; it employs Adaptive Optics (AO) to correct for atmospheric turbulence and a coronagraph for starlight suppression \cite{Soummer_2009, Macintosh_2014, Macintosh_2015, Poyneer_2016}. 

In order to improve the sensitivity of current and future high-contrast instruments, it is critical to understand the factors limiting the performance of current high-contrast systems. Turbulence uncorrected by the AO system has been identified as a major limiting factor for performance \cite{Poyneer_2016, Bailey_2016, Milli_2017}. The residual turbulence produces speckles in the point spread function (PSF), which reduce the instrument's sensitivity. \cite{Racine_1999, Perrin_2003}. Speckle behavior depends on the turbulence source. The effects of the high-altitude jet stream and dome turbulence have been studied extensively \cite{Madurowicz_2018, Cantalloube_2018, Milli_2018}. This work explores the contribution from a different source: mirror seeing.

Turbulence can be parameterized in several ways. The amount of power in phase variations in each spatial or temporal Fourier mode, called the power spectral density (PSD), is one metric. The PSD of atmospheric turbulence is generally assumed to follow the Von Karman spectrum \cite{Karman_1938}; for simplicity, we assume a scale-free Kolmogorov spectrum:\cite{Fried_1966,Hardy_1998,Tatarskii_1961, Roddier_1981}
\begin{equation}
PSD=10^{\alpha} f^\beta
\end{equation}
where $f$ are Fourier modes, $10^{\alpha}$ is the amplitude of the PSD, and $\beta$ is the power-law index of the PSD. We use the following convention to distinguish between spatial and temporal PSD parameters: spatial PSD parameters are denoted with a subscript ``s", while temporal PSD parameters are denoted with a subscript ``t". The rate at which energy is transferred from large scale eddies to smaller scale eddies is embodied in $\beta$. In frozen flow turbulence \cite{Taylor_1938}, $\beta_{s} = -11/3$ and $\beta_{t} = -8/3$ \cite{Conan_1995}. This assumption implies that a single layer of turbulence can be regarded as fixed phase screen propagating horizontally through the atmosphere. The strength and speed of the turbulence is frequently distilled into two parameters:  the coherence length and the coherence time, respectively denoted as $r_0$ and $\tau_0$. The coherence length, also called the Fried parameter, describes the lateral distance across the aperture in which the RMS phase aberration of the incoming wavefront is one radian. The full width at half maximum (FWHM) of the PSF, also called the seeing disk, is $\epsilon_0 \sim \lambda/r_0$ where $\lambda$ is the observing wavelength. The coherence time, $\tau_0 \propto r_0 / v_0$, where $v_0$ is the characteristic velocity of the turbulence. AO systems are generally designed using these turbulence models with assumptions about the local $r_0$ and $\tau_0$ values of the observatory \cite{Madurowicz_2018}. If any of these assumptions are violated, there is a risk of under-designing the system in some aspect, whether it is deformable mirror stroke budget or loop speed. 

 Turbulence introduced in and around the telescope dome can lead to mediocre performance on nights with otherwise exceptional observing conditions. We use the general term ``dome seeing'' to refer to any source of additional turbulence in the dome and ``mirror seeing'' to refer to turbulence sourced specifically from the telescope mirror(s). Studies done with seeing-limited instruments found that the PSF FWHM and Strehl Ratio (the ratio of the peak PSF intensity to that of a perfect Airy pattern) are strongly correlated with temperature differences throughout the dome \cite{Racine_1991, Lowne_1979}. Furthermore, laboratory studies observed that mirror seeing develops as soon as the mirror's temperature exceeds the ambient air temperature \cite{Iye_1991}. Recent results from Robo-AO are also suggestive of the effects of dome seeing on AO instrumentation \cite{Jensen-Clem_2018}. Anecdotally, some GPI data sets have shown poorer performance on nights with otherwise exceptional seeing conditions, as measured by the observatory seeing monitors. This motivates the need for a study of mirror seeing that pertains to high-contrast imaging. Although, we limited our study to a single instrument, these results could also be generalized to other Gemini instruments such as GEMS or ALTAIR and likely to other observatories as well. 

According to the above studies, the key to reducing dome seeing is to control the temperature inside the dome. For the past few years, the temperature control strategy at GS has been to measure the outside air temperature near the start of each night, then use that temperature as the set point for the dome air conditioning for the following afternoon. This strategy performs well when the weather is stable, but it is not ideal when weather fronts cause rapid changes in temperature from one day to the next. The primary mirror (M1) temperature is not actively controlled, and is often considerably warmer than the outside air, as it is a large, heavy structure with a much slower cool-down rate. 

In this work, we investigate the effects of mirror seeing on AO performance, using data from the The Gemini Planet Imager Exoplanet Survey (GPIES)\cite{Nielsen_2019}. GPIES is a direct imaging campaign that began in 2014 to search for exoplanets and circumstellar disks around $\sim$600 young, nearby stars. Each campaign observation is archived together with observatory environmental metadata and AO telemetry. This large, uniform dataset enables us to study the relationship between various environment and telescope parameters with GPI performance. Additionally, we can do a more detailed analysis of mirror turbulence with the AO telemetry data. The datasets used in this study are described in Section \ref{sect:Data}. The filtering and matching of datasets as well as the techniques used to calculate turbulence parameters are described in Section \ref{sect:Methods}. Our results are presented and discussed in Section \ref{sect:Results}. We conclude in Section \ref{sect:Conclusion}. 

\section{Data}
\label{sect:Data}

\subsection{GPIES observations}
A GPIES campaign observation is comprised of 20 -- 40 sequential H-band (1.6~$\mu$m) integral field spectrograph (IFS) images, each with a 60 second exposure. Each exposure is automatically reconstructed into a spectral cube (x, y, $\lambda$), which we refer to as ``raw images.'' A record of the instrument, observatory, and ambient environment states, along with a time stamp and target identification information are saved in the image header. At the end of an observing sequence, individual raw IFS images automatically undergo calibration and PSF-subtraction and are combined into a single ``final'' IFS image by the GPIES data pipeline \cite{Perrin_2014,Wang_2015,Perrin_2016,Ruffio_2017}. 

An important metric used for monitoring GPI's performance is contrast, the minimum detectable planet-to-star flux ratio at a given projected separation from the host star. The 5$\sigma$ contrast at a given separation is defined as five times the local standard deviation of the noise in an annulus centered at that separation. This calculation first takes place for each wavelength slice in a cube, and then the values for the central 50\% of the spectral band are median-combined to produce a single contrast curve that is associated with the observation \cite{Wang_2015,Ruffio_2017}. Typical ``raw contrast'' values at $0.4''$ are $\lesssim 10^{-4}$. The ``final contrast'' is typically 10-50 times better than the raw contrast. 

The raw and final data products are logged in the GPIES database as they are produced \cite{Wang_2017}. This includes the raw and final 5-$\sigma$ contrast measured at $0.2''$, $0.4''$, and $0.8''$ separations. For simplicity, we consider only the raw and final contrast at $0.4''$ separation. At smaller separations performance can be affected by coronagraph misalignment, and at larger separations images may be background limited \cite{Bailey_2016}. In total, we have access to 28,387 raw images and 719 final images, which were gathered between November 2014 and January 2019. 

\subsection{GPI AO telemetry}
GPI's AO system \cite{Poyneer_2016} consists of a Shack-Hartmann wavefront sensor (WFS) and two deformable mirrors (DMs). The WFS uses 2~x~2 pixel quad-cell centroiding to measure the slope of the wavefront across each subaperture in the full telescope aperture, which is 43~subapertures wide. Each subaperture samples 18~cm of the wavefront and these measurements are performed at a rate of 1~kHz on stars brighter than 8th magnitude in I-band, and 500~Hz on fainter stars. The phase of the wavefront is reconstructed in Fourier modes \cite{Poyneer_2002}. Low-order modes are sent to the woofer DM and the tip/tilt stage, high-order modes are sent to the tweeter DM. The realtime control computer estimates the modal coefficients for tip, tilt, and focus and removes them from the signal that drives the control loop. 

Several types of AO data are periodically recorded for analysis purposes, 2--5 times per target, at the discretion of the observer. A full telemetry set contains a record of the WFS measurements, DM and tip/tilt commands, along with any other offsets applied. The datasets used here are 10--22~sec long, and a total number of 2009 AO telemetry sets were recorded between September 2014 and January 2019.

A direct measurement of the AO system's performance is the residual wavefront error ($\sigma_{\rm WFE}$), defined as the error measured by the WFS after applying a correction with the DM. An instantaneous approximation of $\sigma_{\rm WFE}$, based on the RMS subaperture slopes, is saved to the header of each raw IFS frame (See Section \ref{subsect:Filtering Data}). We also estimated the $\sigma_{\rm WFE}$ from reconstructed wavefront telemetry as discussed in Section \ref{subsect:characterizing AO telemetry}. 

\subsection{Gemini observatory data}
\begin{figure}[h]
\begin{center}
\begin{tabular}{c}
\includegraphics[height=7.5cm]{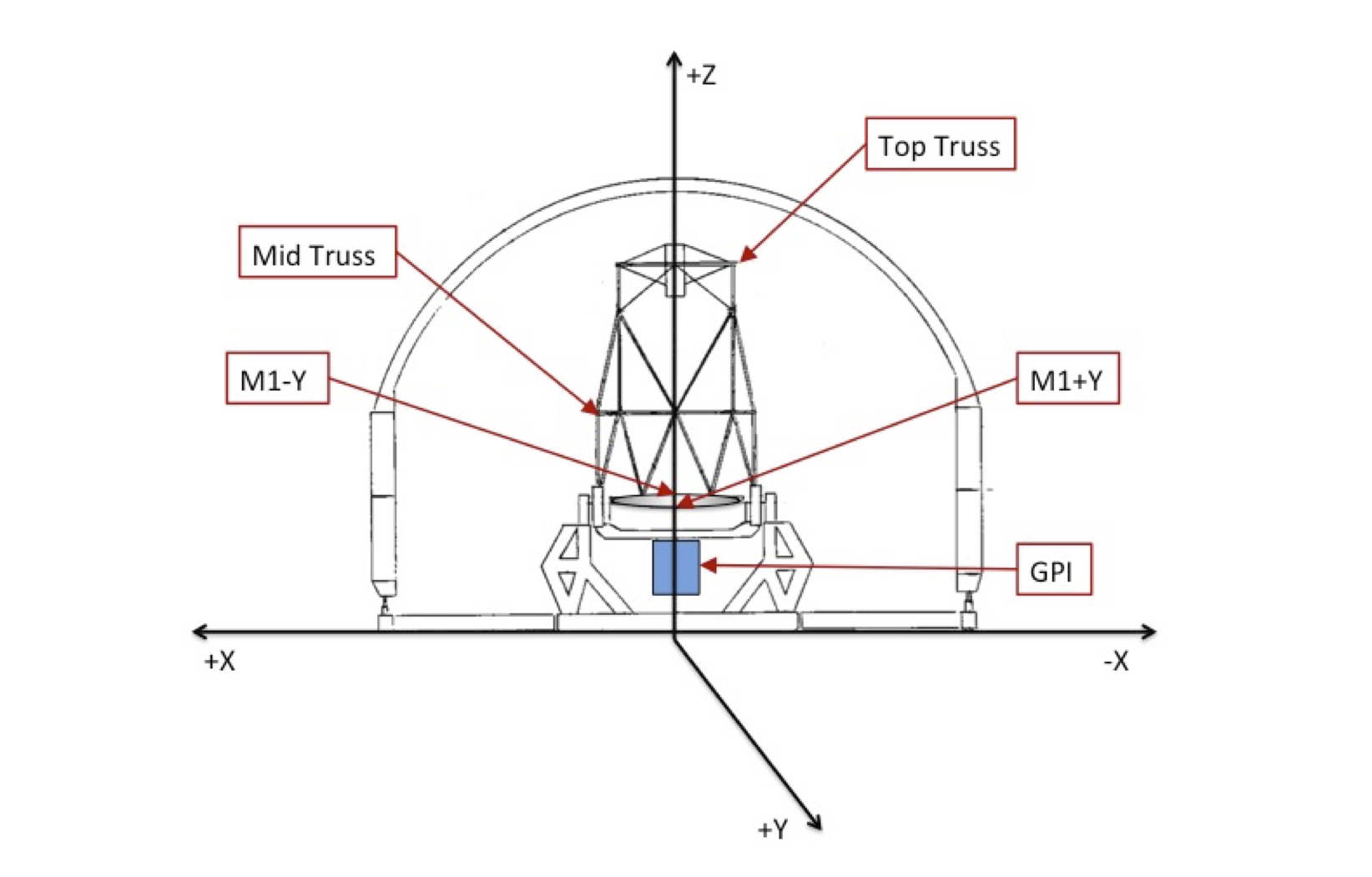}
\end{tabular}
\end{center}
\caption 
{ \label{fig:GS_schematic}
Schematic of Gemini South telescope and observatory dome. Arrows mark the location of temperature sensors.} 
\end{figure}

There are multiple temperature sensors placed in and around the telescope dome at GS. Each sensor records the temperature every five minutes. The approximate location of the interior sensors relevant to this study are displayed in Figure \ref{fig:GS_schematic}. At each location there is one sensor directly attached to the metal truss surface (Minco S651PDX24B \cite{Minco_2019}) and another that is suspended freely in the air next to it (Omega RDT-805 \cite{Omega_2019}). On M1, there are two individual sensors mounted directly on the outer rim. The ``M1-Y'' sensor is attached to the side that tips downwards when the telescope slews to lower elevation, while the ``M1+Y'' sensor is attached to the side that tips upwards. The precision of the temperature data is $\sim1\%$ at 25~${}^{\circ}C$, according to the sensor specifications. Although precise, the sensors may be subject to constant bias offsets, as will be discussed in Section \ref{subsect:Filtering Data}. Additionally, the outside air temperature is recorded by a weather tower adjacent to GS.

The $\epsilon_0$ and $\tau_0$ values at the time of observation were provided by the observatory's Differential Image Motion Monitor (DIMM) and Multi-Aperture Scintillation Sensor (MASS), respectively \cite{Tokovinin_2007}. These instruments are not located in the dome, but rather in a separate tower $\sim$80~m away. Consequently, they are insensitive to mirror or dome seeing, a crucial point for our study.  

\section{Methods}
\label{sect:Methods}

\subsection{Filtering, matching, and calibrating GPI science data and temperature readings}
\label{subsect:Filtering Data}
IFS raw and final datasets were filtered as follows. To ensure high signal-to-noise images and telemetry, we included in our sample only bright stars (I-band magnitude $<$ 7). This also guaranteed that the AO system was operating at 1 kHz for all images in the sample. We also excluded images that were taken in the worst seeing conditions or when there was no recent measurement ($\lesssim40$~min) from the MASS/DIMM. According to the MASS/DIMM, the median seeing recorded at Cerro Pachon from the years 2014 to 2019 was $\epsilon_0 = 0.9''$ and $\tau_0=1.5$~ms. We only kept the images corresponding to the best 85\% seeing conditions:$\epsilon_0 < 1.2''$ and $\tau_0>0.8$~ms. These seeing cuts combined culled almost 80\% of the remaining dataset, in large part because no MASS/DIMM values were available after early 2017. Next, we rejected final images where the field of view (FOV) rotated by less than twice the FWHM of the PSF for a planet orbiting its star at $0.4''$ $(\theta_{\rm FOV}>12.2^{\circ}$)\footnote{$\theta_{\rm FOV}> \frac{2 \lambda}{D\rho}$, where $\lambda = 1.65~\mu m$, $D = 8~m$, and $\rho = 0.4''$.}. The final contrast depends the FOV rotation because the GPI data pipeline relies in part on the angular differential imaging observing technique. Sequences with a lower FOV rotation suffer loss of sensitivity due to self-subtraction during the PSF subtraction process. After all of these cuts, 2,977 raw frames and 120 final images remained. The remaining data represent 57 different observing nights between September 2014 and February 2017. 

Lastly, not all sequences attained the same total integration time. To normalize all values to an equivalent of a full sequence (38~min), we scaled final contrast by $\sqrt{t_i / 38~min}$, where $t_i$ is the total integration time. This scaling was verified to hold after the minimum $\theta_{\rm FOV}$ was achieved, based on a multi-hour c~Eri dataset.

\begin{figure}
\begin{center}
\begin{tabular}{c}
\includegraphics[height=8cm]{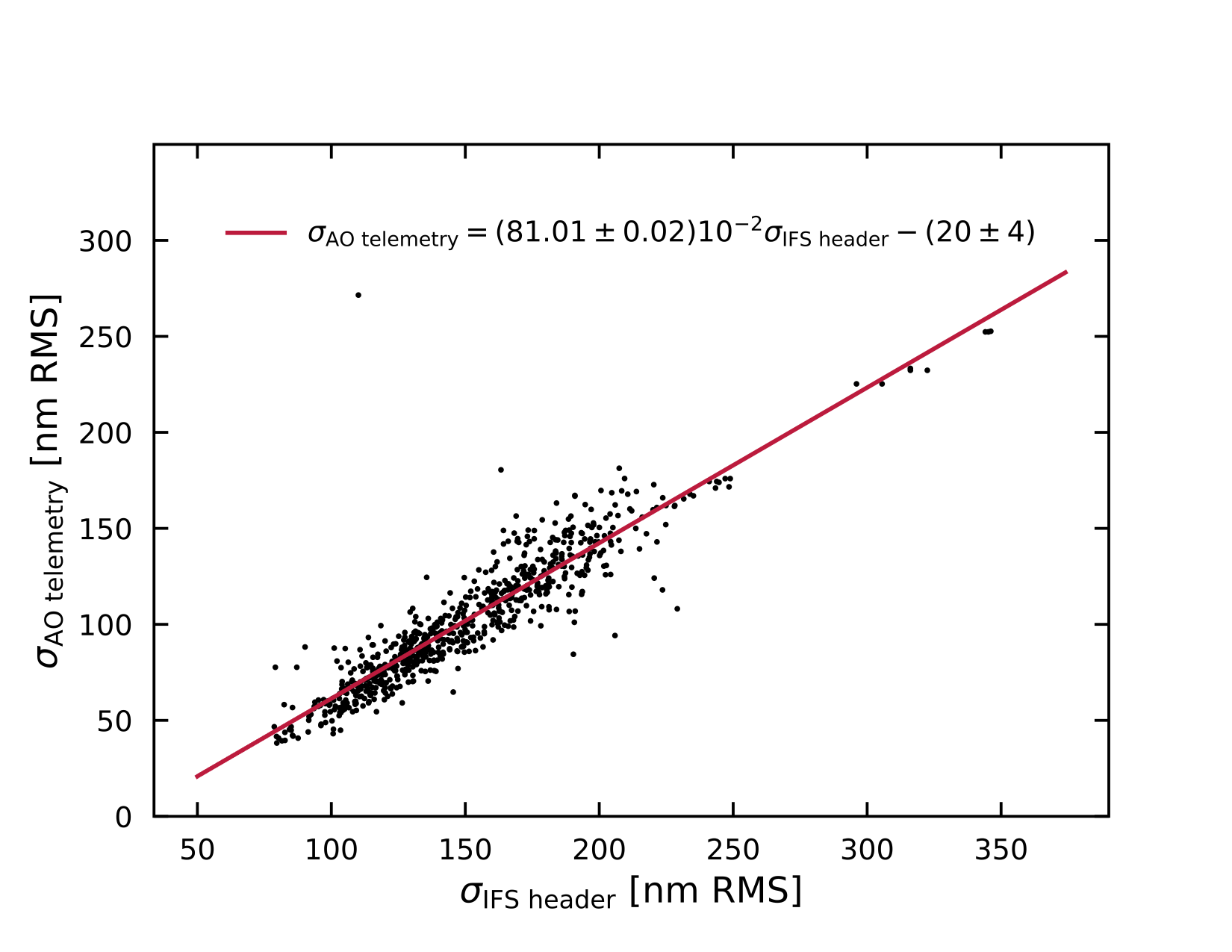}
\end{tabular}
\end{center}
\caption 
{ \label{fig:wfs_calibration}
Comparison of residual wavefront error derived from AO telemetry, $\sigma_{\rm AO~telemetry}$, to the instantaneous approximation saved to the IFS image headers, $\sigma_{\rm IFS~header}$ (black points). The linear fit is overlaid (red line), and was used to correct $\sigma_{\rm IFS~header}$ in subsequent analysis.} 
\end{figure}

The approximate $\sigma_{\rm WFE}$ that is stored in the IFS headers is overestimated because the calibration between the RMS displacement of the WFS centroids and the actual wavefront error was never measured. We obtained the linear relationship: 
\begin{equation}
\sigma_{\rm AO~telemetry} = (81.01 \pm 0.02)10^{-2}~\sigma_{\rm IFS~header} - (20 \pm 4)~[\rm nm~RMS]
\end{equation}
using a subset of images that had contemporaneous residual phase maps reconstructed from full AO telemetry (as described in Section \ref{subsect:characterizing AO telemetry}). The data and fit are displayed in Figure \ref{fig:wfs_calibration}. We corrected all IFS header values using this fit.

We also filtered and calibrated the GS temperature data. Bad or missing data in the GS temperature logs are recorded with a null value; these entries were removed. We cross-checked the M1 temperatures against simultaneous readings from the GPI AO bench temperature sensor and the dome air and truss sensors. Both M1 sensors revealed significant offsets. The observatory staff calibrated these two sensors on August 3, 2017. Unfortunately, the offsets were not recorded during the calibration process. To correct the historical data, we determined the zero-point offset for each sensor in the following way: first, we split the data into before and after calibration sets; from January 1, 2014 to August 2, 2017 and from August 4, 2017 to December 31, 2018, and then folded them by year. Next, we fit a line to the before calibration data from August 4, to December 31 and subtracted the fit from the after calibration data. In the final step, we calculated the DC offset that made the mean of the difference equal to zero. The average offset value for each sensor is displayed in Table \ref{tab:GS_offsets}. The dome air temperature reported by the various truss air temperature sensors also showed moderate systematic offsets. The lower-truss sensor reading most closely matched the temperature reported by the weather tower, while the upper- and mid-truss sensors reported temperatures $1-2~^\circ$C cooler. Because the calibration of these sensors were not independently checked, as M1 sensors were, we chose to use only the readings from the lower-truss sensor. We stress, however, that absolute temperatures were not essential in this study, as we investigated only the relationship between GPI performance and relative temperature changes.

\begin{table}[htb]
\caption{Calculated temperature offsets for GS temperature sensors ($T_{corrected} = T_0$ + offset).} 
\label{tab:GS_offsets}
\begin{center}       
\begin{tabular}{|c|c|} 
\hline
\rule[-1ex]{0pt}{3.5ex}  Sensor & Offset [${}^{\circ}C$] \\
\hline\hline
\rule[-1ex]{0pt}{3.5ex}  M1+Y & 3.5  \\
\hline
\rule[-1ex]{0pt}{3.5ex}  M1-Y & 2.1   \\
\hline
\end{tabular}
\end{center}
\end{table}

Ultimately, a single value of each environment parameter (M1 temperature, dome temperature, $\tau_0$, and $\epsilon_0$ was associated with each raw and final image for all subsequent analysis. We averaged the two corrected M1 sensors to produce an instantaneous M1 temperature. We matched each raw science frame with its closest-in-time temperature readings. If no temperature was recorded within thirty minutes of the observation, we excluded the image from our sample; this removed $\sim25$ of the raw images. Finally, we matched final science frames with the average M1 and dome temperatures and the average seeing recorded during the entire observation sequence. 

\subsection{Characterizing turbulence with AO telemetry}
\label{subsect:characterizing AO telemetry}
We quantified the characteristics of mirror turbulence by examining the spatial and temporal PSDs of the wavefronts at the entrance pupil (ie: before AO correction). In theory it is possible to directly measure mirror seeing with open-loop measurements from the WFS. But in practice, open-loop wavefront errors are too large to be measured by GPI's WFS. Thus, rather than directly measuring the open-loop phase, we reconstructed the ``pseudo open-loop'' phase from closed-loop AO telemetry. A detailed description of this process can be found in Reference \citenum{Poyneer_2009} and in Reference \citenum{Srinath_2015}. The closed-loop residual phase measured by the WFS, $\Phi_{\rm WFS}$, is given by:
\begin{equation}
\label{eq:WFS}
\Phi_{\rm WFS} = \Phi_A - \Phi_{\rm DM},
\end{equation}
where $\Phi_A$ is the pseudo open-loop phase and $\Phi_{\rm DM}$ is the DM commanded phase. Each pseudo open-loop time series was mapped onto a slightly oversized cube containing (48~x~48) subapertures~x~22,000~ms, with a pixel scale of 0.18~m at M1. Any uncontrolled data points, such as those located outside the aperture area, on the aperture edge, or behind the secondary mirror were set to zero. 

Tip and tilt were already removed from the signal by the realtime control computer; however some static aberrations remained. To estimate the low-order static aberrations, we calculated the time-average phase map and fit it with the first six Zernike polynomials, under the assumption that any low-order turbulence-induced wavefront error would average to zero over the 10-22 second time series. We used the python optical simulation package Poppy \cite{Perrin_2016a} to calculate normalized Zernike coefficients.

\begin{figure}[h]
\begin{center}
\begin{tabular}{c}
\includegraphics[width=\textwidth]{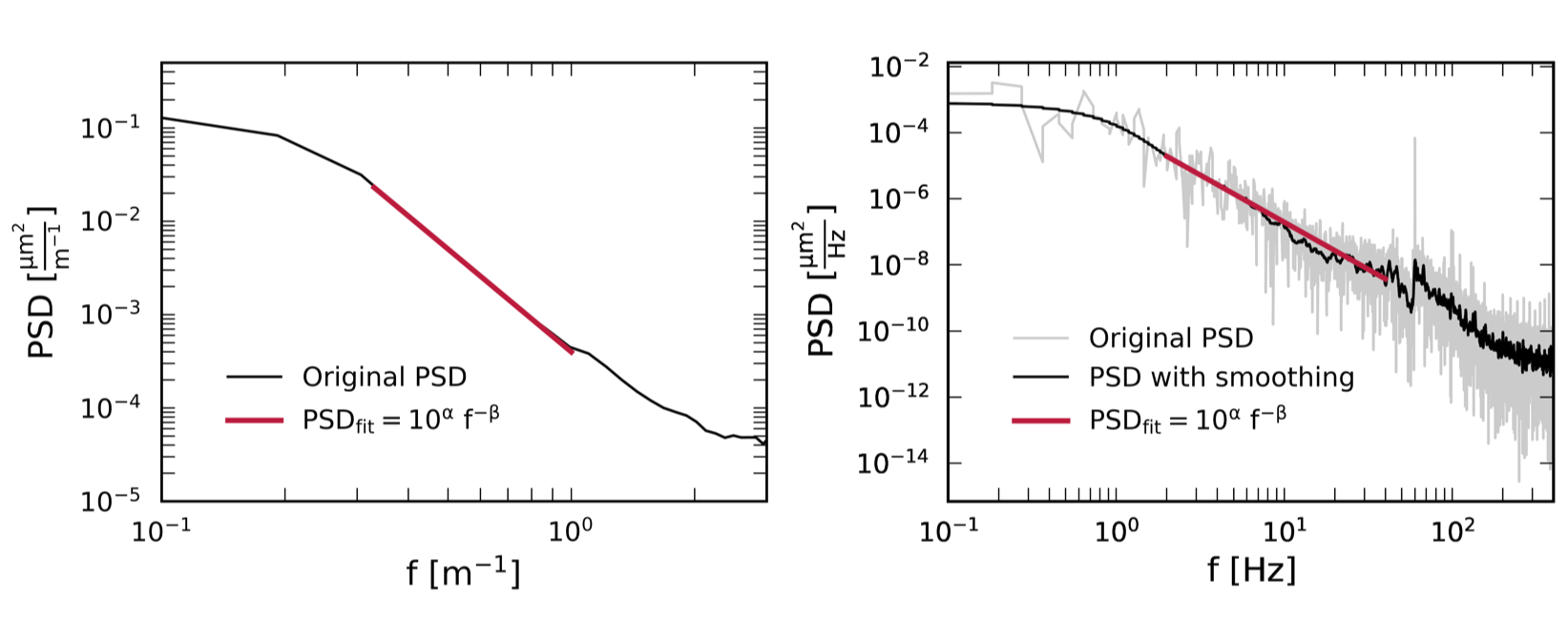}  
\\
(a) \hspace{8cm} (b)
\end{tabular}
\end{center}
\caption 
{ \label{fig:c_Eri_psd}
Spatial (left) and temporal (right) PSDs of pseudo open-loop wavefront errors derived from AO telemetry recorded while observing c Eri on 12/19/2015. We superimposed the temporal PSD obtained after applying the Savitzky-Golay filter (black) with the original temporal PSD (grey).} Power law fits (red line) were performed in the inertial range of the PSDs. 
\end{figure}

We then used the 2D Discrete Fourier Transform (DFT) to decompose the phase signal at each timestep into its Fourier modes. Given the open loop phase $\phi(x,y,t_i)$ and the aperture $a(x,y)$, the modal coefficients at each time step $t_i$ are 
\begin{equation}
\label{eq:sp_DFT}
F_{xy}(k_x,k_y,t_i) = \frac{1}{\sum_{x=0}^{N-1}\sum_{y=0}^{N-1}a(x,y)}
        \sum_{x=0}^{N-1}\sum_{y=0}^{N-1}{\phi(x,y,t_i)}a(x,y)
        e^{-2\pi j(k_x x +k_y y)}
\end{equation}
Computing the square of the spatial modes, yields the spatial PSD, which is the following:
\begin{equation}
\label{eq:sp_PSD}
P_{xy}(k_x,k_y) = \langle| F_{xy}(k_x,k_y,t)|^2\rangle_t
\end{equation}
where $\langle \rangle_t$ is the time-average of the individual PSDs. An example of the spatial PSD (left) and the temporal PSD (right) of the turbulence recorded while observing c Eri on 12/19/2015 is shown in Figure \ref{fig:c_Eri_psd}. 

The length scales captured by GPI range from 0.36~m to 8~m, where 0.36~m is twice one subaperture diameter and 8~m is the full aperture diameter. We fit a power law of the form $\rm PSD=10^{\alpha_s}f^{\beta_s}$ from the middle region of the spectrum $\rm (0.3~m^{-1},1.0~m^{-1})$, which is highlighted in Figure \ref{fig:c_Eri_psd}. This selection ensures that our parameters are free from tilt and outer scale effects at lower spatial frequencies and WFS aliasing effects at higher spatial frequencies.

This process was repeated to estimate the average temporal PSD of the DM actuators:
\begin{equation}
\label{eq:t_PSD}
P_{t}(\omega) = \langle|F_{t}(x,y,t)|^2\rangle_{xy},
\end{equation}
where $\langle \rangle_{xy}$ is the average of all the DM actuators in the aperture and the DFT of each time series is equal to:
\begin{equation}
\label{eq:t_DFT}
F_{t}(\omega) = \frac{1}{N}
        \sum_{t=0}^{N-1}{\phi(x,y,t)}a(x,y)
        e^{-2\pi j(\omega t)}.
\end{equation}
The resulting temporal PSD, which is displayed in Figure \ref{fig:c_Eri_psd} on the right, appears noisier than the spatial PSD because the position of each DM actuator is more finely sampled than the spatial modes. Thus, we used a Savitzky-Golay \cite{Savitzky_1964} filter with a window length of 101~ms and polynomial order of 5 to reduce the noise in the temporal PSD. The Savitzky-Golay filter splits the data into discrete segments and approximates the data inside each interval with a polynomial fit. After we smoothed the PSD, we fit a power law of the form $\rm PSD=10^{\alpha_t}f^{\beta_t}$ from (2~Hz, 40~Hz)to avoid potential biases associated with aliasing and/or vibrations that may be introduced by the telescope.

\begin{figure}[h]
\begin{center}
\begin{tabular}{c}
\includegraphics[width=\textwidth]{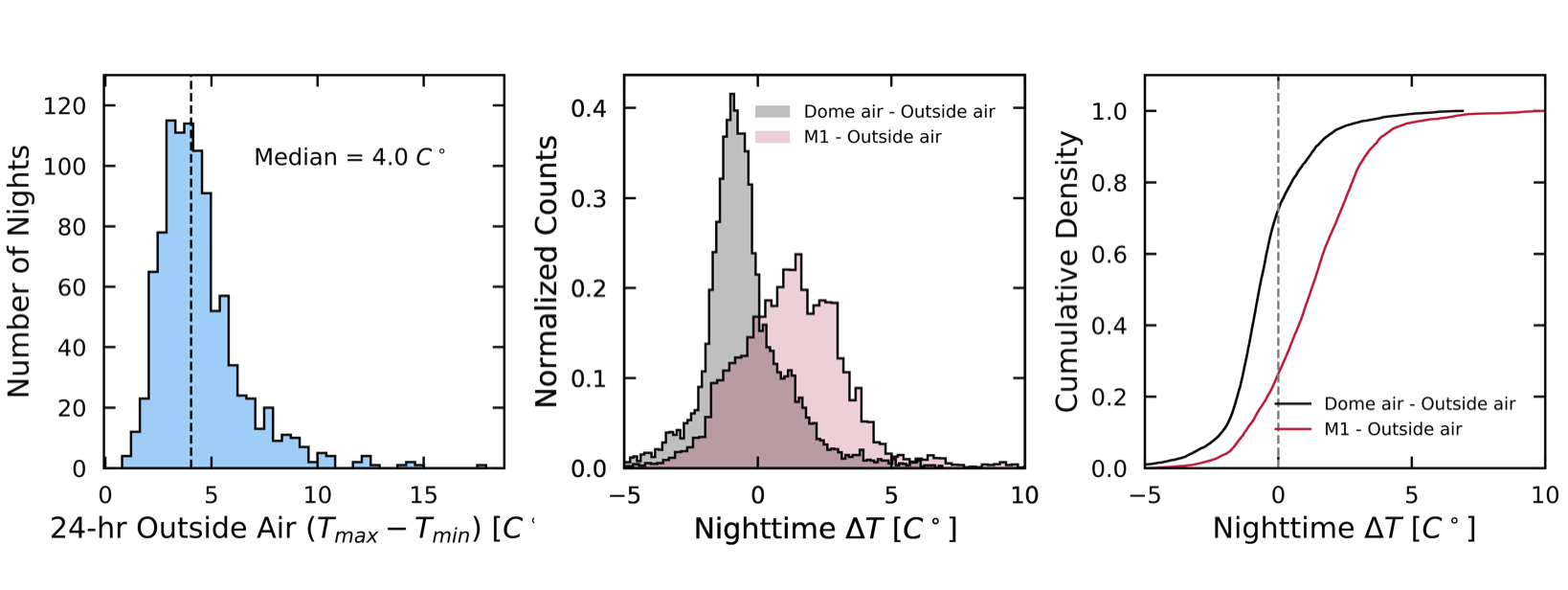}  
\\
(a) \hspace{4.7cm} (b) \hspace{4.7cm} (c)
\end{tabular}
\end{center}
\caption 
{ \label{fig:temp_hist}
Temperatures at the Gemini South observatory. (a) Histogram of the daily min-max range of outside air temperatures. (b) Histogram of instantaneous temperature differences between dome air or M1 and outside air at night, for those nights when GPI was observing, with DC offset corrections as described in Section \ref{subsect:Filtering Data}. (c) Cumulative density plot of the data in (b)}
\end{figure}

\section{Results}
\label{sect:Results}

\subsection{Temperature conditions at Gemini South}
We first investigated the typical temperature conditions at GS using temperature records from the years 2015--2019, which include all nights, not only those with GPI data. We excluded nights with only partial records. We required at least 10 measurements recorded in a 24~hr period, and at least four hours between $T_{\rm max}$ and $T_{\rm min}$. A total of 1001 nights, out of an original 1221, remained. Figure \ref{fig:temp_hist}(a) shows a histogram of $(T_{\rm max}-T_{\rm min})$ of the outside air for each calendar day, split at sunset. The distribution peaks at 4~$^{\circ}C$, but occasionally reaches 15~$^{\circ}C$, illustrating the wide range of day-night temperature swings that the observatory must cope with. We measured the nighttime instantaneous temperature differences between M1, dome air, and ambient air, in order to identify potential sources of dome seeing. We only considered nights in which GPI took on-sky data. We plot their density histograms in figure 4 (b), and their cumulative densities in figure 4 (c). The reported median temperature of the dome is 0.7~$^{\circ}C$ cooler than ambient while the reported median temperature of M1 is 1.2~$^{\circ}C$ warmer than ambient. However, as discussed previously, while the calibration of the M1 sensors has been validated, those of the outside air sensor and the dome air sensor have not. Thus, while the distributions of dome-ambient and M1-ambient difference are correct, the medians could be biased. From the standard deviations of the two distributions (1.6~$^{\circ}C$ and 2.0~$^{\circ}C$ respectively), we can conclude that the dome more quickly tracks ambient temperature than M1.

\begin{figure}[h]
\begin{center}
\begin{tabular}{c}
\includegraphics[width=\textwidth]{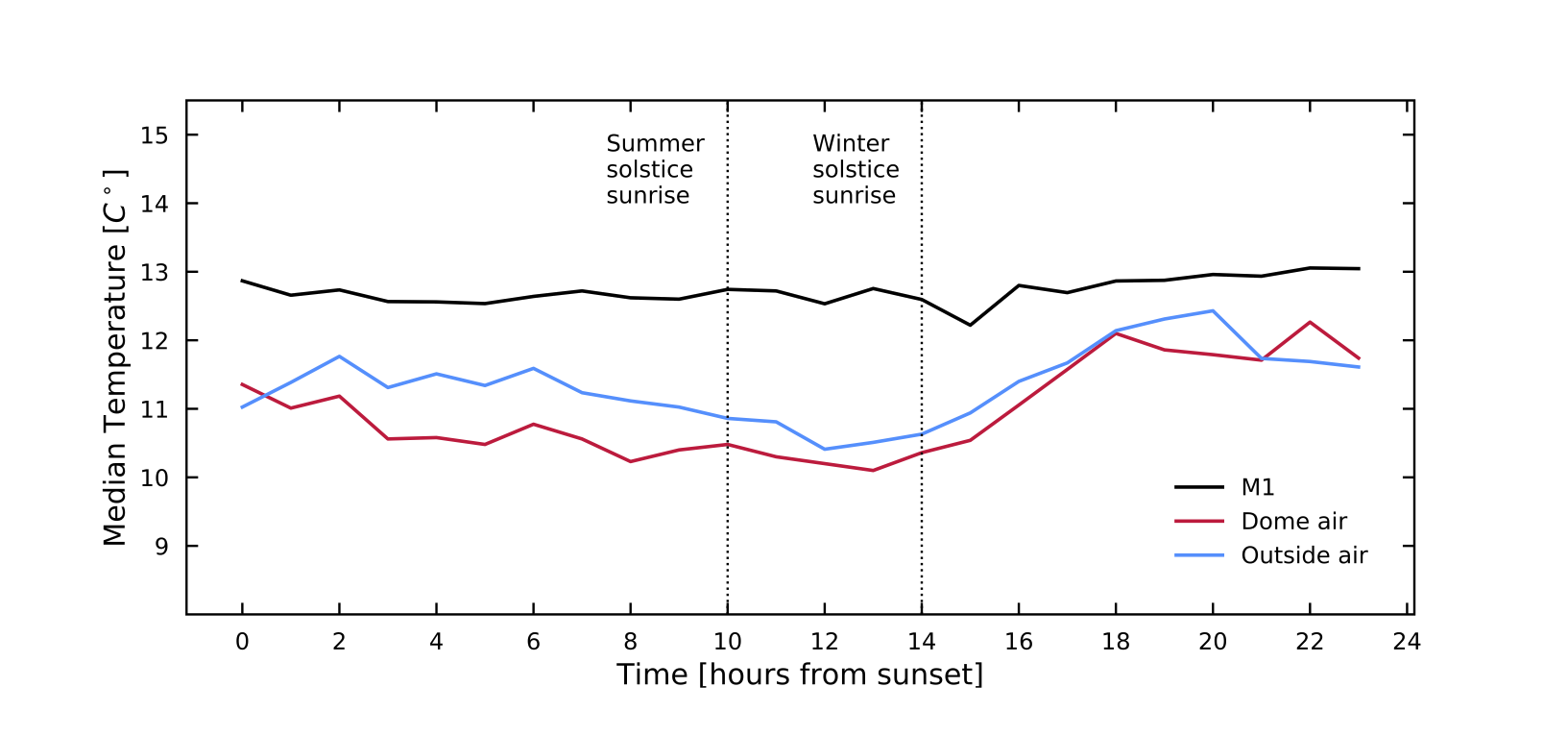}
\end{tabular}
\end{center}
\caption 
{ \label{fig:temp_24hr}
Median 24-hour temperature profile at GS for nights on which GPI took on-sky data. The black, red, and blue lines represent the median temperature of M1, dome air, and the outside air.
Vertical dashed lines mark the approximate length of the shortest and longest nights of the year.} 
\end{figure}

Figure \ref{fig:temp_24hr} shows the median 24-hour temperature profile of M1, dome air, and outside air for nights in which GPI took on-sky data. It shows that the min--to--max range of M1 is small compared to the min--to--max range of the ambient air. We postulate that M1 typically remains warmer than ambient air during the night, resulting in additional ``mirror seeing'' and in GPI AO performance degradation. The exact thermal balance of any observatory is complex. M1 may never be in good equilibrium after sudden jumps in the ambient temperature, which on Cerro Pachon are frequently $>~4~^{\circ}C$ in a day. There could also be potential sources of heat such as electronics in or near the mirror cell. This is an ongoing area of study at the Gemini Observatory.

\subsection{Effect of temperature differentials on residual wavefront error and contrast}

\begin{figure}[h]
\begin{center}
\begin{tabular}{c}
\includegraphics[height=8cm]{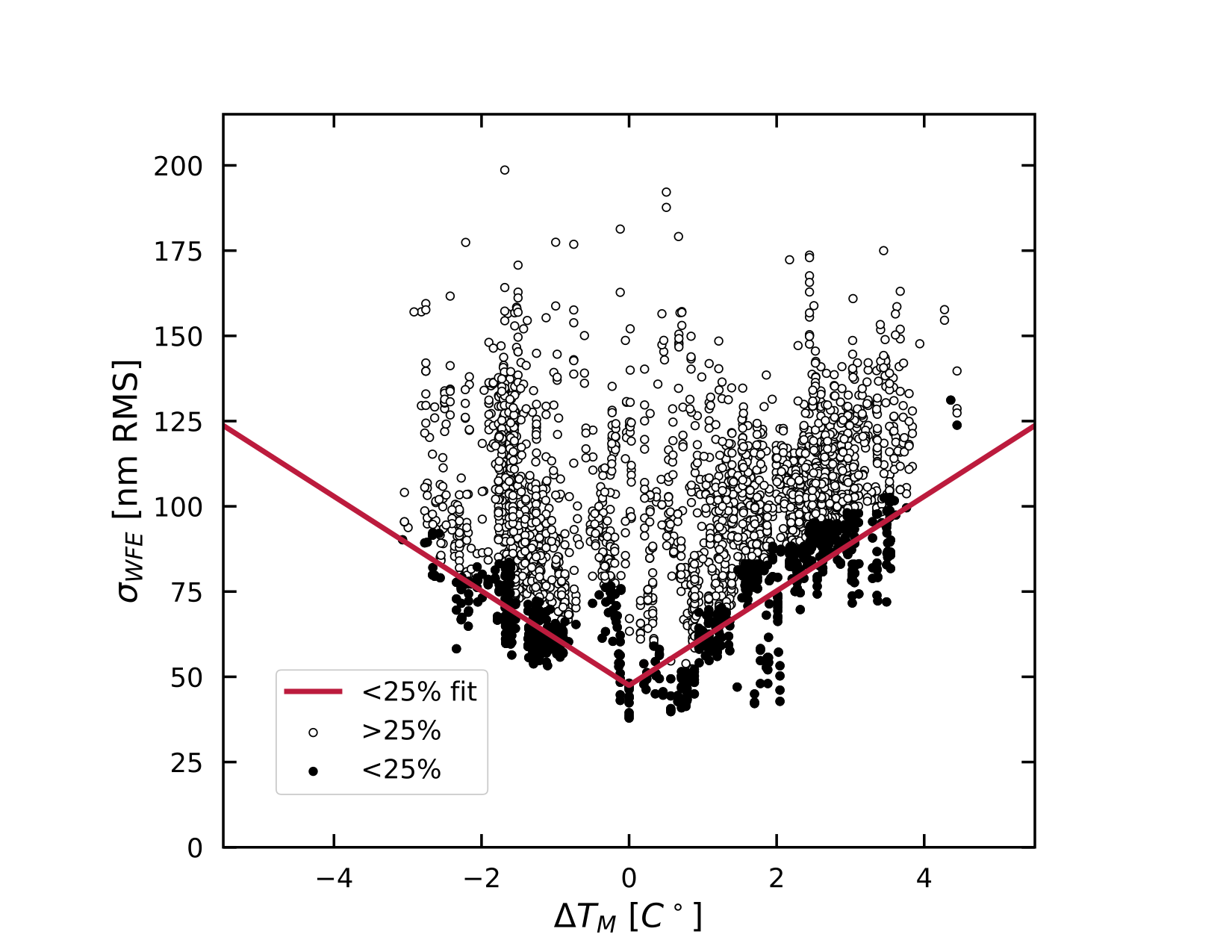}
\end{tabular}
\end{center}
\caption 
{ \label{fig:WFE_vs_delT}
$\sigma_{\rm WFE}$  vs. $\Delta T_M$. The temperature difference between M1 and ambient sets the lower limit on $\sigma_{\rm WFE}$. The red line indicates a fit of the bottom 25$^{th}$ percentile of the data, which are the points shaded in black.} 
\end{figure}

Next we explored the correlation between the instantaneous $\sigma_{\rm WFE}$ recorded in the GPI image headers and the temperature differences between M1 and the outside air. The result, which is displayed in Figure \ref{fig:WFE_vs_delT}, shows that $\sigma_{WFE}$ becomes larger as the temperature difference between M1 and the outside air increases. Some vertical scatter is present due to additional contributions from other sources affecting performance, such as stellar magnitude or atmospheric seeing conditions. With appropriate calibration, measurements from the observatory seeing monitors could potentially be used to correct for some of the atmospheric seeing contribution; this analysis is left to future work. Nonetheless, the performance limit imposed by mirror seeing is clearly captured by the lower envelope of the data, which corresponds to the best seeing conditions. The model\cite{Tallis_2018} we used to capture the relationship between $\sigma_{WFE}$ and $\Delta{T}_M$ is:

\begin{equation}
\label{eq:wfe_delT_fit}
\sigma_{WFE} = a |\Delta{T}_M| + b
\end{equation}
where $a$ and $b$ are free parameters. To isolate the impact of $\Delta{T}_M$, we only fit the bottom 25$^{th}$ percentile of each 0.5~$^{\circ}C$ temperature difference bin. The resulting fit is:
\begin{equation}
\label{eq:wfe_t}
\begin{aligned}
\sigma_{WFE} {} & = (13.9 \pm 0.4)~|\Delta{T}_M| + (47.5 \pm 0.7)~[\rm nm~RMS]
\end{aligned}
\end{equation}
Our fit predicts that we obtain an additional 41.6 $\pm$ 1.2~nm of wavefront error when M1 is $3^\circ$C warmer than the outside air. This behavior is consistent with results from previous studies of mirror seeing\cite{Racine_1991,Lowne_1979,Iye_1991} when $\Delta{T}_M > 0^\circ$C. Past studies observed a weaker effect at $\Delta{T}_M<0^\circ$C. Our data are consistent with an equal slope at both warmer and cooler temperatures, although hints of a weaker effect at cooler temperatures are present. We note that our data span a larger range of negative temperatures than most previous studies.

\begin{figure}[h]
\begin{center}
\begin{tabular}{c}
\includegraphics[width=\textwidth]{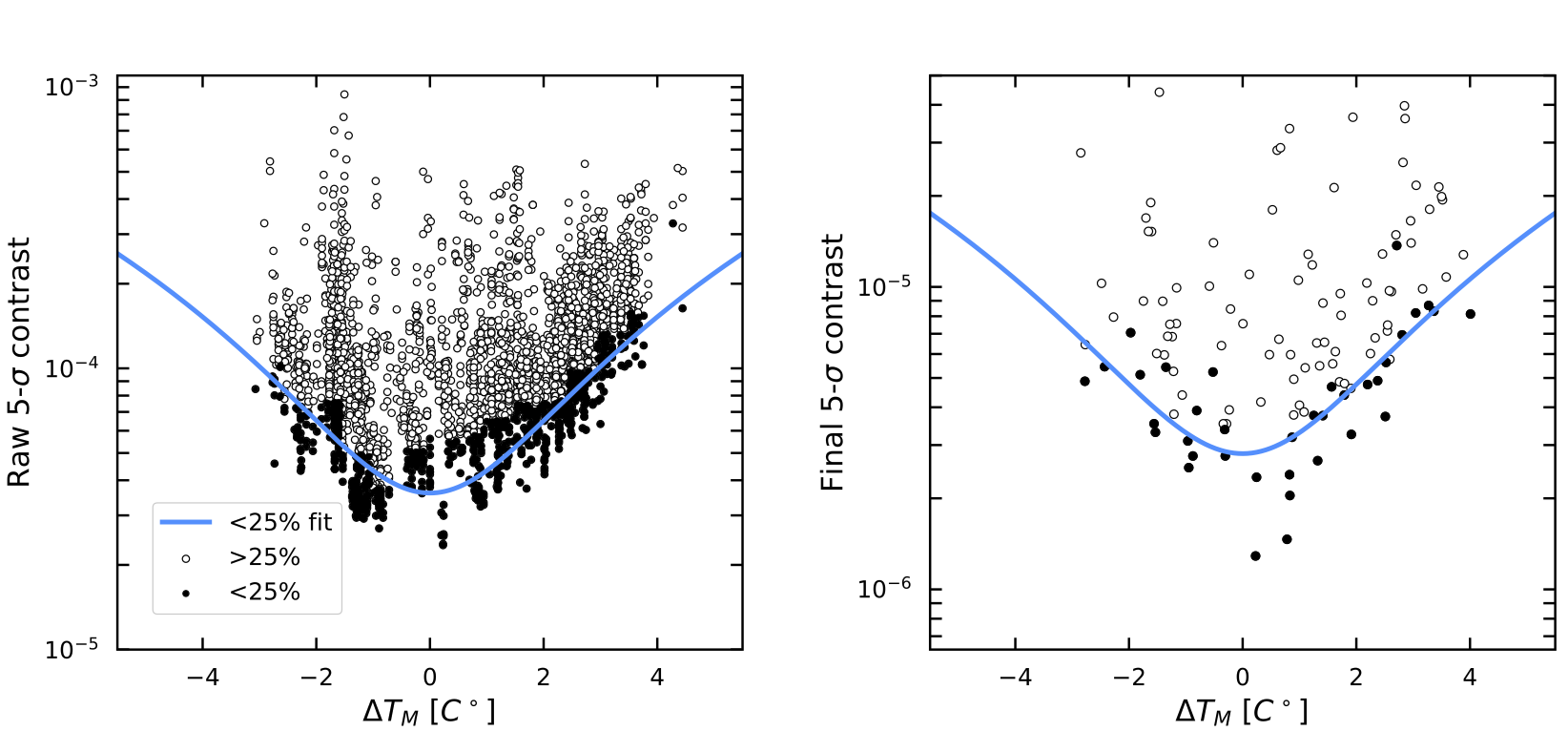}  
\\
(a) \hspace{8.25cm} (b)
\end{tabular}
\end{center}
\caption 
{ \label{fig:cont_vs_temp}
Image contrast vs. $\Delta T_M$. (a) is raw contrast while (b) is final contrast. Blue line is the fit of the best 25$^{th}$ percentile of the data (black points). On nights when the temperature of the mirror diverged by more than 3~$^{\circ}C$ from the outside air temperature, both raw and final contrast degraded by $\sim 3~\times$.} 
\end{figure}

We also quantified the effects of mirror seeing on contrast performance, which is the quantity of interest when imaging planets. Previous correlation studies of GPI performance showed that raw contrast and final contrast both scaled with $\sigma_{\rm WFE}^2$ at close-to-intermediate separations \cite{Poyneer_2016, Bailey_2016}. Therefore, we expected that raw contrast and final contrast to have the following relationship:
\begin{equation}
\label{eq:cont_delT_rel}
Contrast~at~0.4'' = c~\Delta{T}_M^2 + d
\end{equation}
where $c$ and $d$ are free parameters. Figure \ref{fig:cont_vs_temp} displays raw contrast (left) and final contrast (right) plotted as a function of $\Delta T_M$. Once again we see that the temperature difference sets a floor to the achievable contrast. For both raw and final contrast, we again fit only the bottom 25$^{th}$ percentile of each 0.5~$^{\circ}C$ temperature difference bin. The fit for raw contrast is:
\begin{equation}
\label{eq:raw_cont_fit}
\begin{aligned}
\rm  Contrast~at~0.4'' = (7.3 \pm 0.1) 10^{-6}~\Delta{T}_M^2 + (3.6 \pm 0.1)10^{-5},
\end{aligned}
\end{equation} 
and the fit for final contrast is:
\begin{equation}
\label{eq:f_cont_fit}
\begin{aligned}
\rm Contrast~at~0.4'' = (3.4 \pm 0.6) 10^{-7}~\Delta{T}_M^2 + (1.4 \pm 0.3)10^{-6}.
\end{aligned}
\end{equation}
Our fits show that when M1 is 3~$^{\circ}C$ warmer than the outside air, both raw and contrast degrade by a factor of $3 \pm 1$. This is an apparent discrepancy with Reference \citenum{Xuan_2018} results for the Keck/NIRC2 Vortex Coronagraph instrument performance, which did not find a relationship between achievable contrast and temperature differences. However, an overwhelming majority of their data had $\Delta{T}_M <$~0.5~$^{\circ}C$, where we see only a very weak effect. Furthermore, Reference \citenum{Xuan_2018} fit all of their data, not only the best 25$^{th}$ percentile. We believe these two factors account for the apparent discrepancy.

\subsection{Effect of temperature differentials on turbulence power spectra}

Evidently high-contrast imaging performance is reduced significantly when M1 is not in thermal equilibrium; however this relationship alone cannot identify the underlying cause. As discussed in Section \ref{sect:intro}, GPI's AO system was designed assuming Kolmogorov turbulence models of the local atmosphere. Perhaps M1 is an extra source of turbulence, resulting in turbulence conditions for which the AO system is no longer optimized. The GS seeing monitors, located outside the dome, cannot directly measure mirror turbulence. However, GPI's well-characterized AO system may be able to. In this section, we check for relationships between the structure of the spatial and temporal PSDs and $\Delta T_M$ measured during the entire GPIES campaign. The spatial PSD is calculated with all temporal frequencies, and we fit only spatial frequencies between $\rm 0.3~m^{-1}$ and $\rm 1.0~m^{-1}$. The temporal PSD is calculated using all spatial frequencies, and we fit only temporal frequencies between 2~hz and 40~hz. Therefore, keep in mind that we cannot rule out effects at other length scales or frequencies. After the data selection outlined in Section \ref{subsect:Filtering Data}, a total of 582 AO telemetry sets remained. We then averaged the parameters that were extracted from the fits for all telemetry sets taken while integrating on the same star, leaving 122 measurements.

\begin{figure}[h]
\begin{center}
\begin{tabular}{c}
\includegraphics[width=\textwidth]{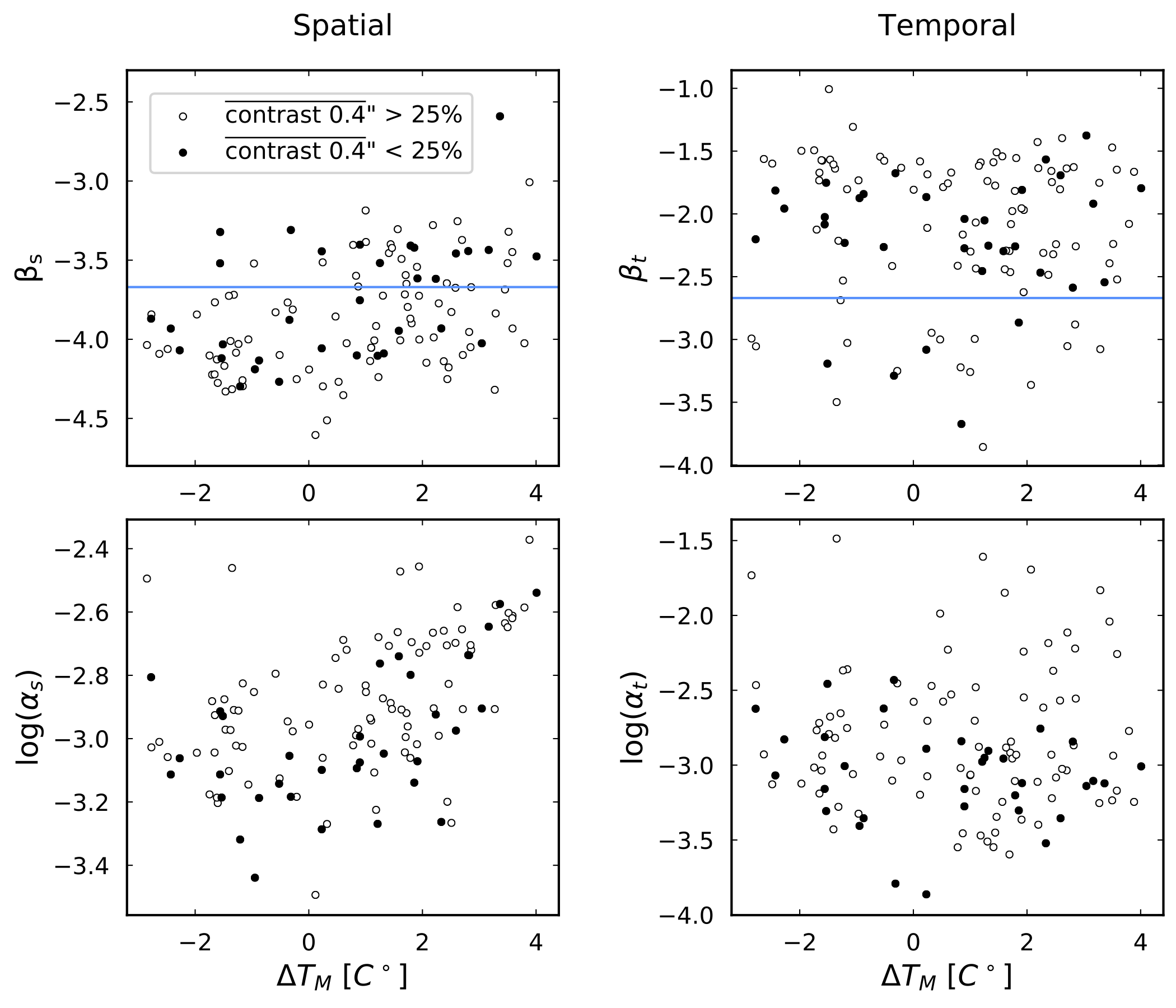}  
\\
\end{tabular}
\end{center}
\caption 
{ \label{fig:psd_slope_corr}
Power law index ($\beta$) and log(amplitude ($\alpha$)) derived from fits of spatial and temporal PSDs are plotted as a function of $\Delta T_{\rm M}$. The points are the mean of each campaign sequence. The black points highlight those datasets with the best 25$^{th}$ percentile contrast, as defined in Figure \ref{fig:cont_vs_temp}. The power law expected for spatial Kolmogorov turbulence (-3.67) and for temporal Kolmogorov turbulence (-2.67) is marked with a line. Only the strength of the turbulence as measured by the spatial PSD shows clear dependence on $\Delta T_{\rm M}$.} 
\end{figure}

 The parameters extracted from our fits, log($\alpha$) and ($\beta$), are plotted as a function of $\Delta T_{\rm M}$ in Figure \ref{fig:psd_slope_corr}. We find that there is no clear dependence between $\beta$ and $\Delta T_{\rm M}$. The $\beta_{s}$ data are uniformly scattered on the expected power-law for spatial Kolmogorov turbulence ($-11/3$), indicating that mirror seeing does not exhibit significant deviations from Kolmogorov behavior at these length scales. The $\beta_{t}$ data also exhibits uniform behavior, but at a slightly higher value than the expected power-law for temporal Kolmogorov turbulence ($-8/3$). This shows that the temporal power spectrum is flatter than what is expected from Kolmogorov, implying an excess of high temporal frequencies with respect to simply translating a Kolmogorov phase screen. Disentangling the relative contributions of noise vs. small-spatial-scale boiling is left to future work.

Now we focus on $\alpha$, which is directly related to the total seeing\cite{Fetick_2019}. We expect the total seeing to increase as the mirror temperature difference increases, and indeed we find a positive trend between $\alpha_{s}$ and $\Delta T_{\rm M}$, albeit at low signal to noise. This, combined with the flat dependence of $\beta_s$ on $\Delta T_{\rm M}$ tells us that the power spectrum is shifted vertically upwards when $\Delta T_{\rm M}$ increases, implying that the intensity of turbulence gets increased at all length scales in the inertial range. Surprisingly, this behavior is less obvious with $\alpha_{t}$, but we hypothesize that our inability to probe smaller length scales and/or other noise sources could be obscuring the trend.

\section{Conclusion}
\label{sect:Conclusion}

In this analysis we studied the effect of the temperature difference ($\Delta T_{\rm M}$) between the primary mirror (M1) and the outside air, a proxy for ``mirror seeing,'' on the performance of the Gemini Planet Imager at the Gemini South Observatory. Our analysis included 2,977 60~sec single exposures, 120 fully reduced observing sequences, and 582 AO telemetry sets recorded between September 2014 and February 2017 during the GPIES campaign, as well contemporaneous temperature and atmospheric seeing measurements. We found that M1 is often warmer than ambient temperature.

We demonstrated that GPI's sensitivity to planets is directly impacted by these temperature differences. Indeed, the best single image (``raw'') contrast and full sequence (``final'') contrast performance occurred when $\Delta T_{\rm M}$ was closest to $0~^{\circ}C$. The raw and final image contrasts degrade by a factor 3 $\pm$ 1 when $\Delta T_M=$~3~$^{\circ}C$. We also analyzed the spatial and temporal PSDs derived from AO system telemetry by fitting a power-law to the inertial region of the spectrum ($\rm 0.3~m^{-1}$ and $\rm 1.0~m^{-1}$) and (2~Hz and 40~Hz). We found that ``mirror seeing'' increased the intensity of turbulence at all length scales in the inertial range, as evidenced by the increase in the spatial PSD amplitude ($\alpha_s$) with $\Delta T_{\rm M}$. We also found that in these frequencies "mirror seeing" was neither more nor less Kolmogorov than turbulence measured when M1 was in equilibrium with the ambient air. Lastly, all datasets showed some deviations from 1-layer frozen-flow expectations; the root cause of which is left to future work.   

We argue that improvements to the GS dome air and mirror temperature control strategies could improve performance of GPI and other adaptive optics instruments. A possible remediation could involve actively cooling M1 during the day with cooling plates behind the glass, which are capable of lowering M1's temperature by $1{}^{\circ}$C/hr \cite{Greenhalgh_1994}. Alternatively, some studies have shown that slowly blowing a stream of cold air above the mirror surface may reduce mirror seeing effects \cite{Lowne_1979}, although this technique has not been tested in high-contrast imaging applications. The observatory staff at GS are continuing to investigate the issue and are in the process of installing a cross-primary laser system that will measure M1 seeing directly.

Our results are generalizable to general-purpose AO systems. The factor of 3 contrast degradation that GPI sees when $\Delta T_{\rm M} = 3~^\circ$C roughly corresponds to a factor of $\sqrt{3}$ increase in wavefront error. Telescope temperature control is critical for maximizing the performance of any AO instrument.

\section{Acknowledgements}
The GPI project has been supported by Gemini Observatory, which is operated by AURA, Inc., under a cooperative agreement with the NSF on behalf of the Gemini partnership: the NSF (USA), the National Research Council (Canada), CONICYT (Chile), the Australian Research Council (Australia), MCTI (Brazil) and MINCYT (Argentina). Additionally, portions of this work were performed under the auspices of the U.S. Department of Energy by Lawrence Livermore National Laboratory under Contract DE-AC52-07NA27344. Vanessa P. Bailey acknowledges government sponsorship; this research was carried out in part at the Jet Propulsion Laboratory, California Institute of Technology, under a contract with the National Aeronautics and Space Administration. 

We thank the referees for their helpful comments and we thank the following people for their contributions to this project; Alex Madurowicz, Varun Harbola, Claire Hebert, Adam Snyder, Eric Nielsen, Jerome Maire, Lea A. Hirsch, and Robert De Rosa. 

\bibliography{report} 
\bibliographystyle{spiebib} 

\end{document}